\documentstyle[vsolj01,graphicx,natbib]{article}

\RequirePackage[T1]{fontenc}

\def\cite{\citealt}

\begin{document}

\title{Orbital and spin periods of the candidate white dwarf pulsar}
\vskip -2mm
\title{ASASSN-V J205543.90$+$240033.5}

\author{Taichi Kato$^1$, Franz-Josef Hambsch$^{2,3,4}$,
        Elena~P. Pavlenko$^5$, Aleksei~A. Sosnovskij$^5$}
\author{$^1$ Department of Astronomy, Kyoto University,
       Sakyo-ku, Kyoto 606-8502, Japan}
\email{tkato@kusastro.kyoto-u.ac.jp}
\author{$^2$ Groupe Europ\'een d'Observations Stellaires (GEOS),
     23 Parc de Levesville, 28300 Bailleau l'Ev\^eque, France}
\author{$^3$ Bundesdeutsche Arbeitsgemeinschaft f\"ur Ver\"anderliche
     Sterne (BAV), Munsterdamm 90, 12169 Berlin, Germany}
\author{$^4$ Vereniging Voor Sterrenkunde (VVS), Oostmeers 122 C,
     8000 Brugge, Belgium}
\author{$^5$ Federal State Budget Scientific Institution
     ``Crimean Astrophysical Observatory of RAS'',}
\author{Nauchny, 298409, Republic of Crimea}

\begin{abstract}
ASASSN-V J205543.90$+$240033.5 has been suggested to be
a white dwarf pulsar by Kato (2021, arXiv:2108.09060).
We obtained time-resolved photometry and identified
the orbital and spin periods to be 0.523490(1)~d and
0.00678591(1)~d = 9.77~min, respectively.
These values strengthen the similarity of this object
with AR Sco.  We estimated that the strength of
the spin pulse is 3.6 times smaller than in AR Sco.
\end{abstract}

\section{Introduction}

   ASASSN-V J205543.90$+$240033.5 is a variable object detected
by the All-Sky Automated Survey for Supernovae (ASAS-SN,
\cite{ASASSN}, \cite{koc17ASASSNLC}).
This object was also independently listed as a candidate
RR Lyr star by \citet{ses17PS1RR} using the Panoramic Survey
Telescope and Rapid Response System (Pan-STARRS1, \cite{PS1}).
It was also independently listed as a variable star
ATO J313.9329$+$24.0092 by \citet{hei18ATLASvar}
using the Asteroid Terrestrial-impact Last Alert System (ATLAS,
\cite{ATLAS}) data.  In \citet{kat21j205543}, one of the authors
(TK) reported superimposed long and short periods and suggested
that this object could be a white dwarf pulsar
similar to AR Sco (\cite{mar16arsco}; \cite{sti18arsco}).

\section{Observation and true orbital period}

   Following this suggestion, FJH reported time-resolved photometry
on seven consecutive night between 2021 August 24 and 30
using a 50-cm telescope located in San Pedro de Atacama (Chile).
The total number of observations was 979 and
the time resolution was 96~s.
EP and AS also supplied a light curve using the 2.6-meter reflector
(Shajn Telescope, ZTSh) in Crimea.  EP suggested that
the orbital period would be longer than the 0.2~d one
suggested by TK in vsnet-alert 26186.\footnote{
  $<$http://ooruri.kusastro.kyoto-u.ac.jp/mailarchive/vsnet-alert/$26186>$.
}

\begin{figure*}
  \begin{center}
    \includegraphics[width=16cm]{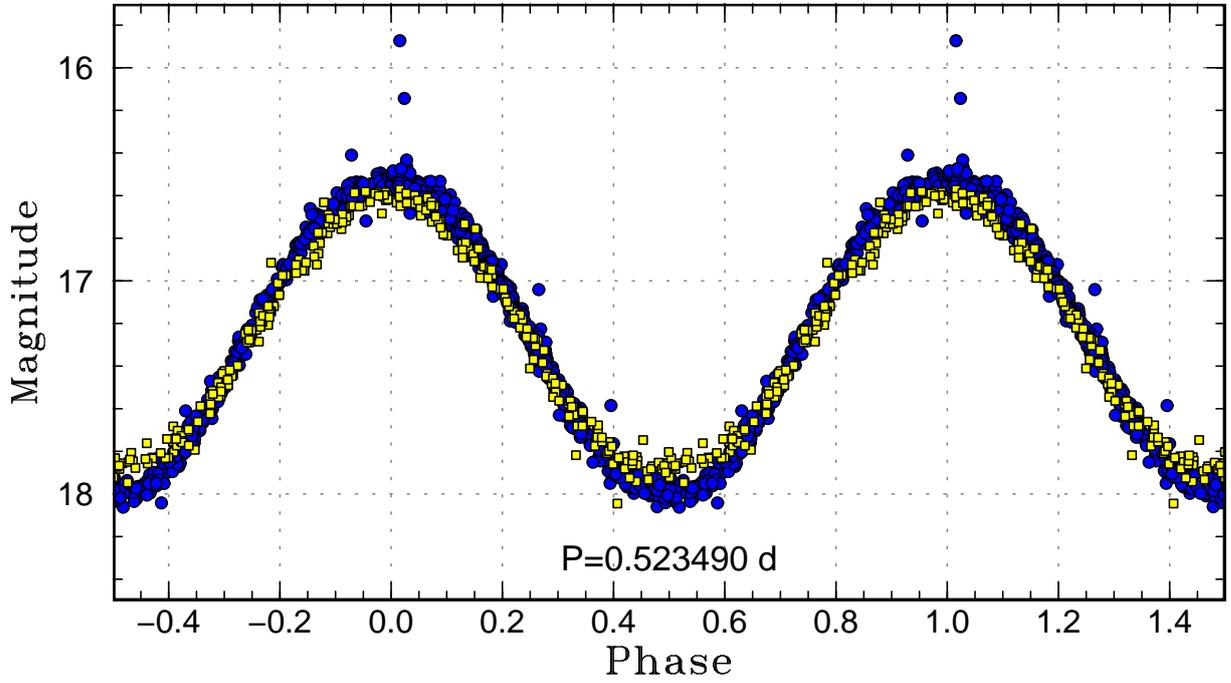}
  \end{center}
  \caption{Phase-folded variations of ASASSN-V J205543.90$+$240033.5
  in the ZTF data.  Filled circles and squares represent
  ZTF $r$ and $g$ observation, respectively.
  The epoch was chosen as BJD 2458773.24.}
  \label{fig:phorb2}
\end{figure*}

\begin{figure*}
  \begin{center}
    \includegraphics[width=16cm]{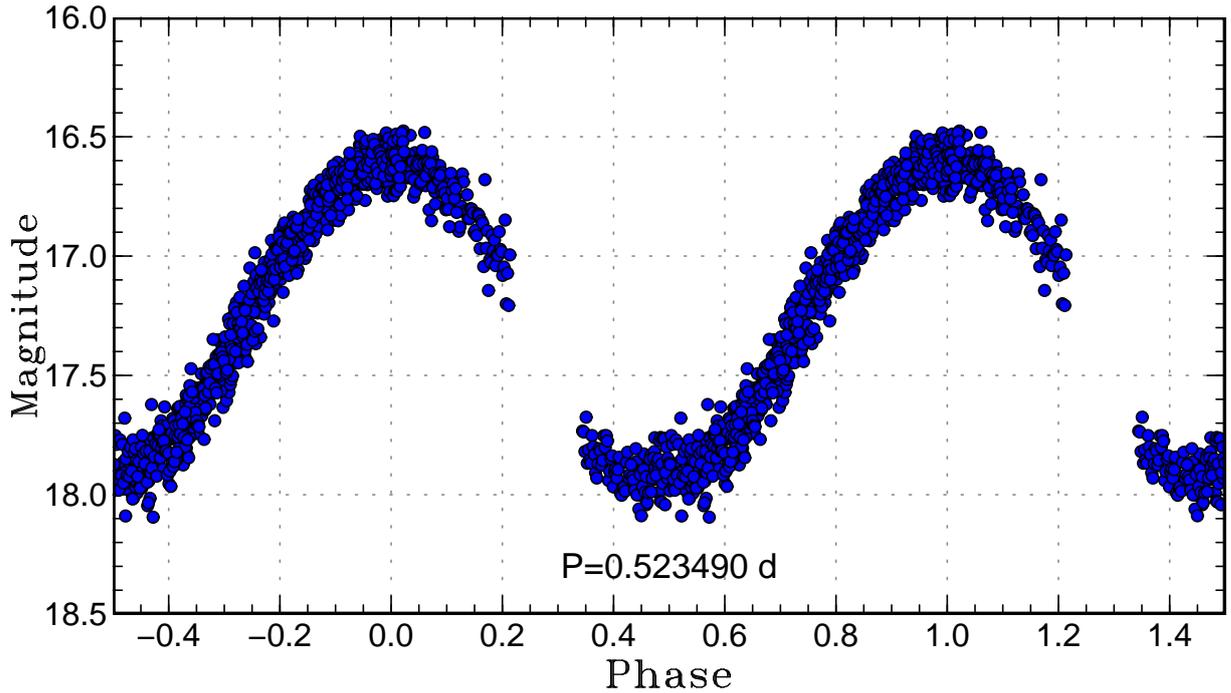}
  \end{center}
  \caption{Phase-folded light curve of ASASSN-V J205543.90$+$240033.5
  using time-resolved photometry by FJH.
  The period and epoch are the same as in figure \ref{fig:phorb2}.
  }
  \label{fig:phorbham}
\end{figure*}

   We reanalyzed the data from Public Data Release 6 of
the Zwicky Transient Facility \citep{ZTF}
observations\footnote{
   The ZTF data can be obtained from IRSA
$<$https://irsa.ipac.caltech.edu/Missions/ztf.html$>$
using the interface
$<$https://irsa.ipac.caltech.edu/docs/program\_interface/ztf\_api.html$>$
or using a wrapper of the above IRSA API
$<$https://github.com/MickaelRigault/ztfquery$>$.
} and found that the true orbital period is 0.523490(1)~d
(first announced in vsnet-alert 26206\footnote{
  $<$http://ooruri.kusastro.kyoto-u.ac.jp/mailarchive/vsnet-alert/$26206>$.
}), not 10.803(2)~d as reported in \citet{kat21j205543}.
The period and error were determined by
the Phase Dispersion Minimization (PDM; \cite{PDM}) method
and the methods in \citet{fer89error} and \citet{Pdot2}.
Figure \ref{fig:phorb2} shows the phase-folded light curve
using the true orbital period.  The light curve is almost
perfectly sinusoidal and is interpreted as the reflection-type
variation caused by strong irradiation from the white dwarf.
This period is confirmed to express all the observations
by FJH (figure \ref{fig:phorbham}).  The initially obtained
period of 10.803~d was an alias of the 0.523490~d period
resulting from the intervals of the nightly ZTF observations.

\section{Spin period}

\begin{figure*}
  \begin{center}
    \includegraphics[width=13cm]{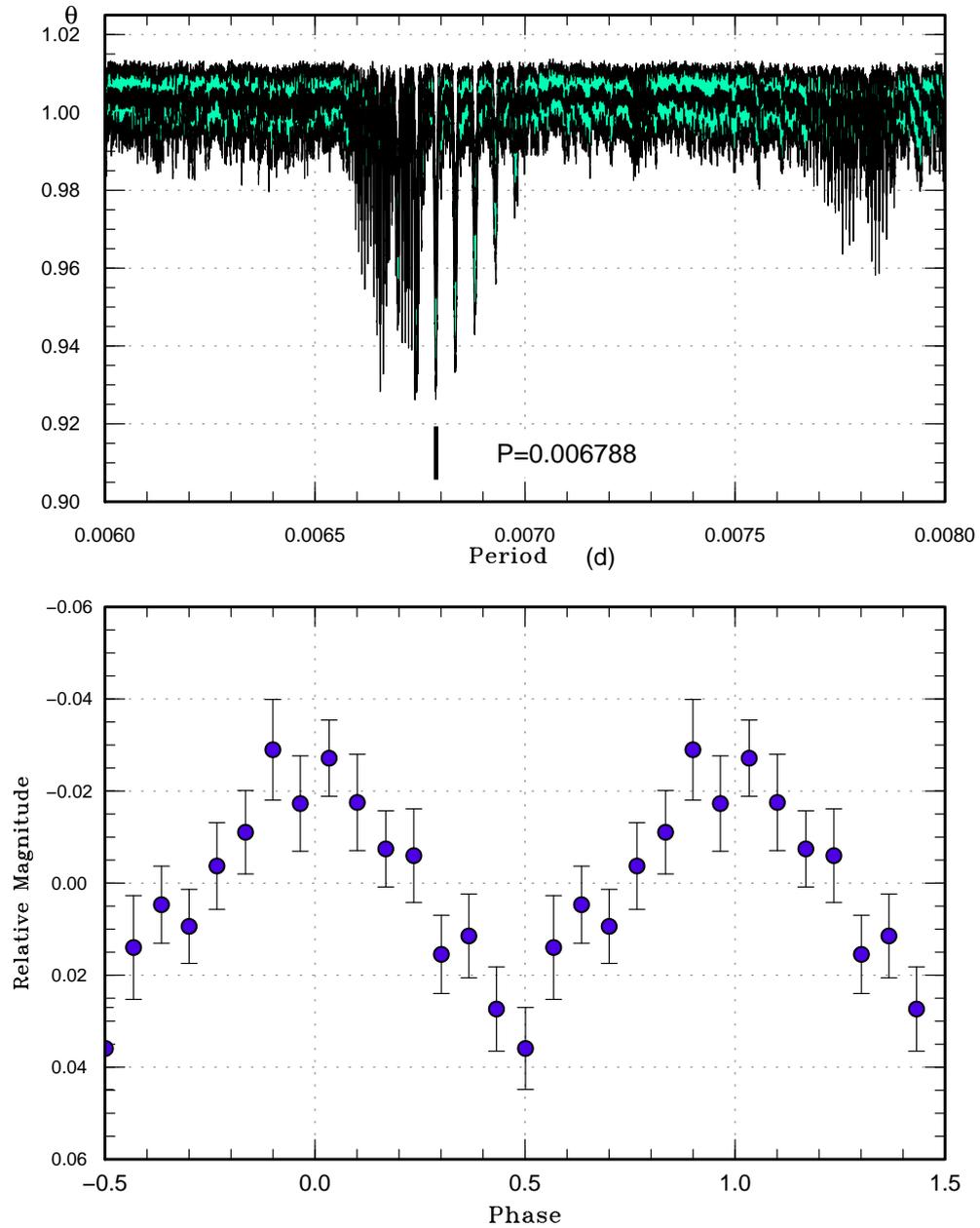}
  \end{center}
  \caption{Pulse profile of ASASSN-V J205543.90$+$240033.5
     using FJH data after removing the orbital variation.
     (Upper): PDM analysis.  We analyzed 100 samples which
     randomly contain 50\% of observations, and performed
     the PDM analysis for these samples.
     The bootstrap result is shown as a form of 90\% confidence
     intervals in the resultant PDM $\theta$ statistics.
     (Lower): Phase-averaged profile.}
  \label{fig:shortham}
\end{figure*}

\begin{figure*}
  \begin{center}
    \includegraphics[width=13cm]{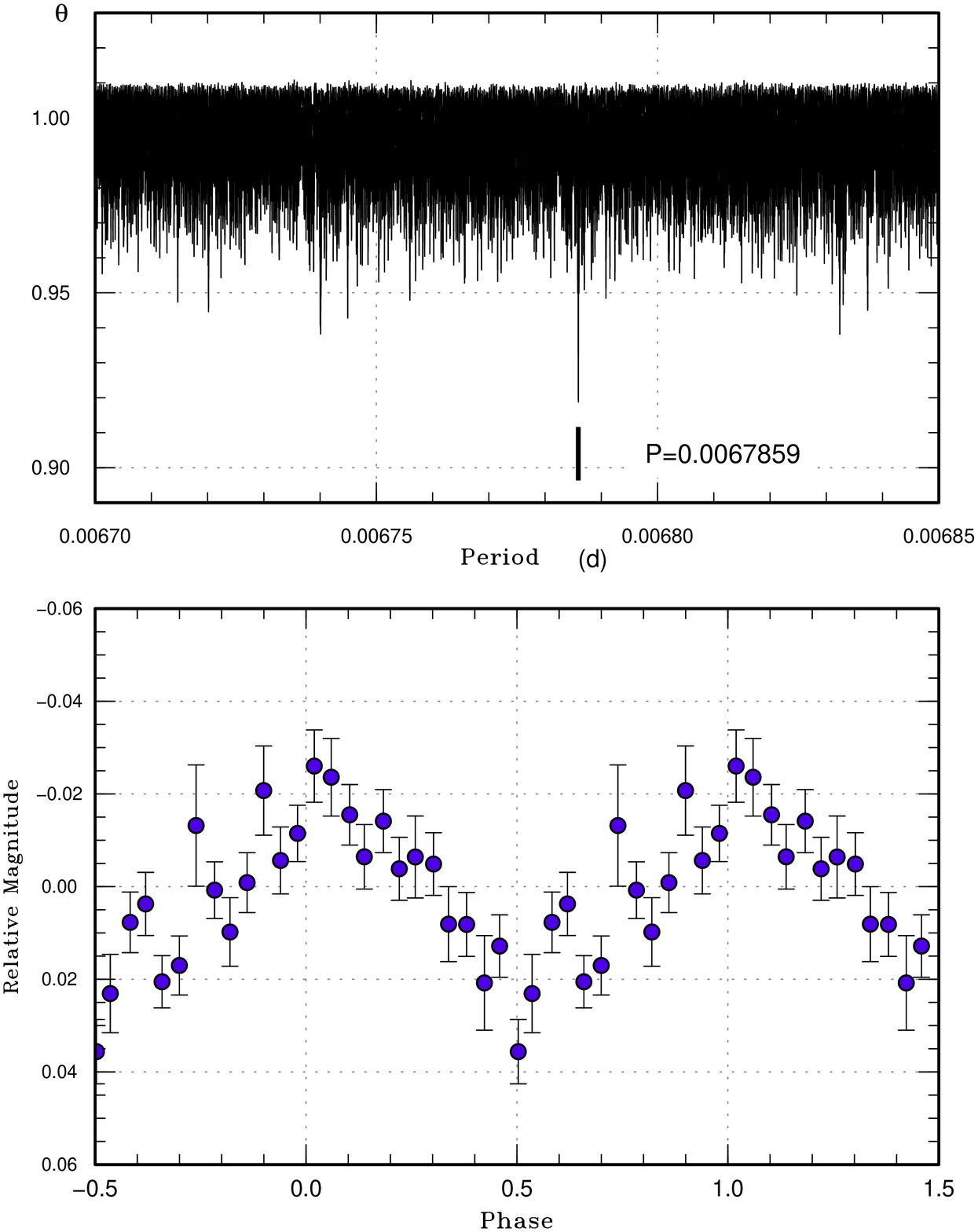}
  \end{center}
  \caption{Pulse profile of ASASSN-V J205543.90$+$240033.5
     using the ZTF after removing the orbital variation.
     (Upper): PDM analysis.
     (Lower): Phase-averaged profile.}
  \label{fig:shortztf}
\end{figure*}

   ASASSN-V J205543.90$+$240033.5 showed short-period
coherent variation in two ZTF time-resolved runs
\citep{kat21j205543}.  The periods on the two nights
were 0.00675(3)~d and 0.00682(2)~d.
The pulse profile using FJH data is shown in
figure \ref{fig:shortham}.  Almost sinusoidal 0.06-mag
variations were clearly detected.  The best period
using this data set is 0.0067882(7)~d, although there
remains a possibility of aliases (such as 0.006742~d).
Using this value, we have been able to identify
the spin period using the ZTF data (figure \ref{fig:shortztf}).
The resultant period is 0.00678591(1)~d = 9.77~min.

\section{Comparison with AR Sco}

   With the newly determined orbital and spin periods,
the similarity between ASASSN-V J205543.90$+$240033.5
has become more evident (table \ref{tab:par},
the periods for AR Sco were taken from \cite{sti18arsco}).
As seen from the almost zero color index in
figure \ref{fig:phorb2}, ASASSN-V J205543.90$+$240033.5 is not
very significantly reddened.  The Gaia color
$G_{\rm BP}-G_{\rm RP}$ is $+$0.38 \citep{GaiaEDR3}.
Considering that the white dwarf and illuminated hemisphere
of the secondary are the dominant source of light, we consider
the unreddened color to be around 0.
On the other hand, AR Sco is more strongly reddened.
The Gaia color $G_{\rm BP}-G_{\rm RP}$ is $+$1.38.  Assuming that
the unreddened colors of these two object are around 0,
we can estimate extinctions by using the formula
$A(G) = 1.89 E(G_{\rm BP}-G_{\rm RP})$ \citep{wan19extinction}.
We adopted $A(G)$ values for 0.72 and 2.61 for 
ASASSN-V J205543.90$+$240033.5 and AR Sco, respectively.
Using Gaia parallaxes \citep{GaiaEDR3}, the unreddened
absolute magnitudes ($M_G$) of these two objects are
$+$5.4 and $+$7.0, respectively.  This indicates that
ASASSN-V J205543.90$+$240033.5 is intrinsically 4.4 times
more luminous than AR Sco, which appears to reflect
the difference in the orbital period.
The mean pulse amplitude is 0.7~mag (90\%) in AR Sco
(figure 10 in \cite{sti18arsco}).  Considering that
the system is 4.4 times more luminous in
ASASSN-V J205543.90$+$240033.5,
the pulse amplitudes (0.06~mag) in the latter corresponds
to 25\% pulse in AR Sco, which means that the strength of
the spin pulse is 3.6 times smaller than in AR Sco.
Whether this difference reflects the difference in
the emission mechanism needs to be studied by further
multiwavelength observations.

\begin{table*}[]
\caption{Comparison between ASASSN-V J205543.90$+$240033.5 and AR Sco}
\label{tab:par}
\begin{center}
\begin{tabular}{lll}
\hline
            & ASASSN-V J205543.90$+$240033.5 & AR Sco \\
\hline
Orbital period (d) & 0.523490 & 0.14853528 \\
Spin period (d) & 0.00678591 & 0.00136804584 \\
\hline
\end{tabular}
\end{center}
\end{table*}

\section*{Acknowledgments}

The author is grateful to Naoto Kojiguchi for supplying
a wrapper code for obtaining the ZTF data.
The author is also grateful to Yusuke Tampo for
processing the VSNET campaign data.

This work was supported by JSPS KAKENHI Grant Number 21K03616.

Based on observations obtained with the Samuel Oschin 48-inch
Telescope at the Palomar Observatory as part of
the Zwicky Transient Facility project. ZTF is supported by
the National Science Foundation under Grant No. AST-1440341
and a collaboration including Caltech, IPAC, 
the Weizmann Institute for Science, the Oskar Klein Center
at Stockholm University, the University of Maryland,
the University of Washington, Deutsches Elektronen-Synchrotron
and Humboldt University, Los Alamos National Laboratories, 
the TANGO Consortium of Taiwan, the University of 
Wisconsin at Milwaukee, and Lawrence Berkeley National Laboratories.
Operations are conducted by COO, IPAC, and UW.

The ztfquery code was funded by the European Research Council
(ERC) under the European Union's Horizon 2020 research and 
innovation programme (grant agreement n$^{\circ}$759194
-- USNAC, PI: Rigault).

\end{document}